\documentclass[preprints,article,accept,moreauthors,pdftex,10pt,a4paper]{Definitions/mdpi}
\firstpage{1} 
\pubvolume{xx}
\issuenum{1}
\articlenumber{5}
\pubyear{2018}
\copyrightyear{2018}
\history{Received: \today}%; Accepted: date; Published: date}
\usepackage{amssymb}
\Title{Millimeter-wave Monitoring of Active Galactic Nuclei with the Africa Millimetre Telescope}

\Author{Michael Backes $^{1,2,\ast}$\orcidA{}, Markus Böttcher $^{2}$\orcidB{} and Heino Falcke $^{3,4,5,6}$\orcidC{}}

\AuthorNames{Michael Backes, Markus Böttcher and Heino Falcke}

\address{%
	$^{1}$ \quad University of Namibia, Department of Physics, Private~Bag~13301, Windhoek, 10005, Namibia\\
	$^{2}$ \quad North-West University, Centre for Space Research, Private~Bag~X6001, Potchefstroom, 2520, South Africa\\
	$^{3}$ \quad Radboud University, Department of Astrophysics/Institute for Mathematics, Astrophysics and Particle Physics~(IMAPP), P.O.~Box~9010, 6525~AJ Nijmegen, The Netherlands\\
	$^{4}$ \quad Max-Planck-Institut für Radioastronomie, Auf dem Hügel~69, 53121 Bonn, Germany\\
	$^{5}$ \quad ASTRON --- Netherlands Institute for Radio Astronomy, Dwingeloo, The Netherlands\\
	$^{6}$ \quad NIKHEF --- National Institute for Nuclear and High Energy Physics, Amsterdam, The Netherlands
}

\corres{Correspondence: mbackes@unam.na}

\abstract{Active Galactic Nuclei are the dominant sources of gamma~rays outside our Galaxy and also candidates for being the source of ultra-high energy cosmic~rays. In addition to being emitters of broad-band non-thermal radiation throughout the electromagnetic spectrum, their emission is highly variable on timescales from years to minutes. Hence, high-cadence monitoring observations are needed to understand their emission mechanisms. The Africa Millimetre Telescope is planned to be the first mm-wave radio telescope on the African continent and one of few in the Southern hemisphere. Further to contributing to the global mm-VLBI observations with the Event Horizon Telescope, substantial amounts of observation time will be available for monitoring observations of Active Galactic Nuclei. Here we review the scientific scope of the Africa Millimetre Telescope for monitoring of Active Galactic Nuclei at mm-wavelengths.
}

\begin{document}
%%%%%%%%%%%%%%%%%%%%%%%%%%%%%%%%%%%%%%%%%%
%
\section{Introduction}
Active Galactic Nuclei~(AGN) and more particularly the subclass of blazars have been a major topic of research throughout the electromagnetic spectrum for the past fifty years. Despite significant progress in understanding the blazar phenomenon~\cite{Boettcher-review}, many open questions remain. Several apparent differences could be attributed to their non-spherical structure and the orientation of the highly relativistic jets relative to the line of sight~\cite{agn:unification}. Still, the underlying particle acceleration and emission processes remain a matter of current research. Connected to the more than one hundred years of quest for the sources of the highest-energetic cosmic rays, hadronic emission scenarios are tested, e.g.~\cite{Boettcher:2005, SMM:reimer}, and recent coincident gamma-ray and neutrino detections point in that direction~\cite{coincident-nu-gamma}.
Even though, hadronic emission scenarios are not uncontested, as also the leptonic Synchrotron Self-Compton~(SSC) model~\cite{model:ssc} is highly successful in explaining the broad-band spectral energy distribution~(SED) for blazars, see e.g.~\cite{mkn421:mwl}.
AGN are variable on all time scales: from optical quasi-periods of several years~\cite{periods:OJ287} down to gamma-ray flares lasting only minutes~\cite{Aharonian:2007pks} and even limiting the size of the emission region to be smaller than~20\% of the gravitational radius of the central black hole~\cite{IC310:lightning}. This motivates intensive efforts for monitoring the variability of AGN across the electromagnetic spectrum.

Following recent positive developments of astronomy in Africa in general~\cite{povic:astro-africa} and Namibia in particular~\cite{backes:astro-nam}, the Africa Millimetre Telescope~(AMT) project aims to build a mm-wave radio telescope in Namibia~\cite{AMT}. It will be the first telescope of its kind on the African continent and one of only few in the Southern hemisphere. The proposed site for the AMT is Mt.\,Gamsberg (23.34$^\circ$S, 16.23$^\circ$E) in the Khomas Highlands of Namibia, at a height of 2,347\,m a.s.l. At very similar latitude to ALMA, this site offers exceptional observing capabilities at~3\,mm (100\,GHz) throughout the year as well as strong though seasonal capabilities at~1.3\,mm (230\,GHz) and~0.8\,mm (345\,GHz), for the latter particularly suited during June through August~\cite{AMT}. The AMT will re-purpose the refurbished structure of the SEST telescope~\cite{SEST} and shall (initially) employ receivers for~3.5\,mm and~1.3\,mm (86\,GHz and 230\,GHz), possibly extended to~0.8\,mm (345\,GHz). One of the main scientific drivers for the AMT are observations of the `shadows' of the black holes at the centres of our Milky Way, Sagittarius~A$^\ast$, and of the radio galaxy M~87 at~3.5\,mm and~1.3\,mm within the Event Horizon Telescope~(EHT) network~\cite{EHT:network}. The shadow of the black hole at the centre of M~87 has recently been imaged for the first time by the EHT~\cite{EHT:M87}. Initial simulations for Sagittarius~A$^\ast$ indicate a significant improvement in image quality and, hence, angular resolution by adding the AMT to the EHT~\cite{AMT:sim}. Still, EHT observations will only make use of a small fraction of the available observation time on the AMT, allowing for high-cadence monitoring observations of AGN to be conducted in addition.

%%%%%%%%%%%%%%%%%%%%%%%%%%%%%%%%%%%%%%%%%%
%\section{Materials and Methods}
\section{Scientific Rationale}
%Materials and Methods should be described with sufficient details to allow others to replicate and build on published results. Please note that publication of your manuscript implicates that you must make all materials, data, computer code, and protocols associated with the publication available to readers. Please disclose at the submission stage any restrictions on the availability of materials or information. New methods and protocols should be described in detail while well-established methods can be briefly described and appropriately cited.
%
%Research manuscripts reporting large datasets that are deposited in a publicly available database should specify where the data have been deposited and provide the relevant accession numbers. If the accession numbers have not yet been obtained at the time of submission, please state that they will be provided during review. They must be provided prior to publication.
%
%Interventionary studies involving animals or humans, and other studies require ethical approval must list the authority that provided approval and the corresponding ethical approval code. 
%
Particularly because of their high variability, monitoring of AGN is crucial to obtain a complete picture of their variability patterns and to understand the underlying phenomena in these enigmatic objects. Across all accessible wavelength regimes, monitoring campaigns are being conducted: Besides the all-sky monitoring in X-rays by MAXI, aboard the International Space Station, and \emph{Swift}-BAT, in high energy gamma~rays by \emph{AGILE}-GRID and \emph{Fermi}-LAT and in ultra-high energy gamma~rays by HAWC, targeted monitoring of AGN is conducted particularly successfully by optical telescopes, e.g. SMARTS~\cite{SMARTS}, GASP~\cite{GASP} of the WEBT~\cite{WEBT}, and the Steward Observatory blazar monitoring program~\cite{Steward}, radio telescopes, e.g. OVRO~\cite{OVRO}, UMRAO~\cite{UMRAO}, and Metsähovi~\cite{Metsahovi1, Metsahovi2}, and the imaging atmospheric Cherenkov telescopes~(IACTs) in the very-high energy gamma~rays. VERITAS and MAGIC pursue long-term blazar monitoring~\cite{magic:monitoring1-own, Backes:icrc11-own} and FACT~\cite{fact:mirrors-own, FACT:JINST} was built for the purpose of blazar monitoring~\cite{Bretz:network-idea-own, Backes:telescope-own, Temme:FACTmon}, even with the idea in mind to set up a world-wide network for continuous monitoring~\cite{Backes:network-own}. In the Southern hemisphere, the H.E.S.S. telescopes conduct very-high energy gamma-ray monitoring of AGN~\cite{HESS:2155monitoring, HESS:1510monitoring}. These IACT observations are regularly complemented by optical monitoring in $R_C$- and $B_C$-bands by the robotic KVA~\cite{KVA} and ATOM telescopes~\cite{atom1,atom2}.

It has long been established by very long baseline interferometry~(VLBI) at up to~43\,GHz (7\,mm) that the high energy gamma-ray emission as measured by \emph{CGRO}-EGRET is coincident with the appearance of new VLBI features (`knots')~\cite{egret-vlbi}. Later, this was underpinned with the observation of a gamma-ray flare coincident with a new VLBI knot in the radio~galaxy M~87~\cite{magic:m87}.
In this context, there have been long-standing programs VLBI flux and morphology monitoring of AGN like MOJAVE in the Northern hemisphere at~15\,GHz (2\,cm)~\cite{MOJAVEmon}, TANAMI in the Southern hemisphere at~22\,GHz (1.3\,cm)~\cite{TANAMImon}, and VLBA-BU-BLAZAR in the Northern hemisphere at~43\,GHz (7\,mm)~\cite{BUmon}. Further, single-dish flux monitoring has been conducted at up to~43\,GHz with the 100\,m~Effelsberg telescope within F-GAMMA~\cite{F-GAMMAmon,F-GAMMAmon2} and with APEX at~345\,GHz (0.87\,mm)~\cite{APEXmon}. Polarimetric monitoring at~86\,GHz and~229\,GHz (3.5\,mm and~1.3\,mm) has been conducted in the Northern hemisphere at the IRAM~30\,m telescope within the POLAMI programme~\cite{polami1, polami2, polami3}. Further, multi-wavelengths monitoring campaigns have been organized, like MARMOT~\cite{MARMOTmon}.
A recent review of mm-VLBI observations of AGN is given in~\cite{mmVLBI}, whereas the major initiative of mm-VLBI monitoring of AGN is conducted by the GMVA~\cite{GMVA} at~86\,GHz (3.5\,mm) and the EHT at higher frequencies. For both initiatives, the only telescopes in the Southern hemisphere are (phased) ALMA and APEX (in Chile) and the South Pole Telescope. Whereas the addition of another Southern hemisphere telescope to the EHT is one of the major and obvious scientific drivers of the AMT, complementing the GMVA at~3.5\,mm VLBI and improving the $u$--$v$-coverage of mm-VLBI observations of Southern AGN is a scientific purpose in its own right: With the superior angular resolution of mm-VLBI observations over cm-VLBI the smallest scale structures, closest to the core of AGN can be resolved. Combined with polarisation imaging, the ordering and mean direction of the magnetic field at the base of the jet can be determined to distinguish between models of highly ordered helical magnetic fields or turbulent ones (possibly with standing shocks), e.g.~\cite{Marscher:model, Sironi:reconnection}.

Though correlation studies between~5\,GHz--8\,GHz VLBI flux densities obtained from the Radio Fundamental Catalogue and \emph{Fermi}-LAT measurements show much stronger correlation for the high energy ($>100$\,MeV) measurements than for the very-high energy ($>50$\,GeV) ones~\cite{Fermi-RFC}, preliminary findings comparing ALMA observations at~230\,GHz~(1.3\,mm) of a sample of~77~AGN from the \emph{Fermi}-LAT~3FGL catalogue (above~100\,MeV) show a significantly higher correlation than comparing the \emph{Fermi}-LAT flux to~1.4\,GHz data~\cite{ALMA-Fermi}. This hints at the mm-wave emission of AGN being more strongly connected to the gamma-ray emission than the cm-wave emission. The AMT will for the first time allow for strictly simultaneous observations of mm-wave and very-high energy gamma-ray emission, due to the close location of the AMT to the H.E.S.S. telescopes, which will allow for unprecedented studies of the emission properties of Southern hemisphere AGN.

One of the major drawbacks of cm radio flux density observations of AGN is that the cm emission is likely produced at different regions of the jet and, hence, for consistent modelling of the spectral energy distribution, these measurements have to be ignored because of the opaqueness of the emission region to cm~radio waves due to synchrotron self-absorption. This changes drastically for observations in the mm-regime. Likely for the~3.5\,mm (86\,GHz) emission but certainly for the~1.3\,mm (230\,GHz) one, synchrotron self-absorption is negligible and the innermost regions of the jet can be probed. Such observations could test whether mm-wave and (very-)high energy emission are produced co-spatially. In this case, both, spectral as well as temporal studies would be significantly enriched. Not only would the mm-wave flux be useful for constraining AGN emission models, but also temporal correlation of the respective lightcurves would be expected --- obeying differences in the electron cooling timescales. The electron cooling timescale for synchrotron emission at~230\,GHz is given by
%Zumindest bei 230\,GHz ist man ziemlich sicher schon im optisch dünnen Synchrotron-Bereich der Hochenergie-Emissions-Region(en). Damit hätte man also einen guten Ansatzpunkt zu untersuchen, inwieweit Radio und Hochenergie-Emission in der gleichen Region erzeugt werden. In dem Fall würde man eben korrelierte Variabilität erwarten - modulo Elektronen-Kühlungszeit. Die Kühlungszeitskala für Elektronen, die bei 230\,GHz strahlen, liegt bei 

\begin{equation}
t_c(\text{obs}) \sim 40 B_G^{-3/2} \delta_1^{-1/2} \text{ days,}
\label{eq:t_c}
\end{equation}
where $B_G$ is the strength of the magnetic field in Gauss and $\delta_1$ is the Doppler factor in units of~10. It is rather well established that emission of AGN close to the core is produced in confined volumes with $R\lesssim 10^{16}$\,cm, which are moving at relativistic speed $\beta_{\Gamma} = \sqrt{1-\frac{1}{\Gamma^2}}$, with a bulk Lorentz factor~$\Gamma$, along the jet. The Doppler factor is defined as $\delta \equiv \frac{1}{\Gamma\left(1-\beta_{\Gamma} \cos\theta^{\text{obs}}\right)}$, with $\theta^{\text{obs}}$ being the viewing angle (with respect to the jet axis) in the observer's rest frame.
%wobei $B_G$ das $B$-Feld in Gauss und $\delta_1$ der Doppler-Faktor in Einheiten von 10 sind.
Considering that neither the magnetic field strength nor the Doppler factor are known and taking into account the possibility for Compton-dominated cooling, resulting in shorter cooling timescales than given in \autoref{eq:t_c}, this well motivates monitoring observations with weekly cadence.
For emission of visual light, the cooling timescale is shorter by a factor of~$\sim100$ than given in \autoref{eq:t_c} above. Hence, for co-spatial emission, the optical light curves should lead the radio ones by this difference. Further, if such a time-lag could be established (assuming an independent estimate of the Doppler factor), the magnetic field strength in the emission region could be estimated (as proposed initially for intra-optical time-lags in~\cite{Boettcher:BLLac}). The magnetic field is generally believed to play a central role in the launching and collimation of AGN jets and in the acceleration of relativistic electrons. Thus, probing the magnetic field in the mm-emitting region will aid in clarifying its role in the particle acceleration (magnetic reconnection, e.g.~\cite{Sironi:reconnection}, vs. shocks, e.g.~\cite{Marscher:shocks}, vs. shear layers, e.g.~\cite{shears}) and its potential for collimating the jet.

Even if no correlation with identifiable time lag between mm and optical flux is observed, the mm-wave monitoring observations would be highly useful: In case the mm-wave variability would not be correlated to the optical variability of a given AGN, this would indicate different emission regions. Possibly, molecular and dust emissions could not only dominate the infrared part of the spectral energy distribution but even the mm-wave emission, as observed for some blazars~\cite{Fumagalli:gas}. This, in turn, would drastically help to constrain external photon fields within the AGN as invoked for external-Compton emission models~\cite{model:ec}.

%Für optische Emission wäre das um einen Faktor~$\sim100$ kürzer. Um diese Zeitskala würde man also bei co-spatialer Emission die Radio-Lichtkurven hinter den optischen verzögert sehen. Der Wert der Zeitverögerung (wenn er gemessen werden kann und man unabhängig eine Abschätzung des Doppler-Faktors hat) könnte dann zur Messung des Magnetfeldes benutzt werden (wie z. B. vorgeschlagen für time-lags innerhalb des optischen Bereichs in~\cite{Boettcher:BLLac}).
%
%Umgekehrt würde Radio-Variabilität, die nicht mit dem optischen korreliert ist, darauf hindeuten, dass man hier unterschiedliche Emissions-Regionen hat. Eventuell könnte hier auch schon Staub- und Molekular-Emission im Infraroten dominieren, wie das bei einigen Blazaren schon gemessen worden ist (z. B.~\cite{Fumagalli:gas}). Das würde wiederum helfen, externe Photonenfelder im AGN einzugrenzen, die man dann für das Extern-Compton-Modellieren gebrauchen kann. 

Obviously, for any significant correlation studies of multi-wavelengths lightcurves, high-cadence monitoring observations are needed as will be supplied by the AMT.

\section{Summary}
%Authors should discuss the results and how they can be interpreted in perspective of previous studies and of the working hypotheses. The findings and their implications should be discussed in the broadest context possible. Future research directions may also be highlighted.
Summarizing, there is ample scope for the AMT to have significant impact in the field of (gamma-ray loud) AGN. Particularly, high-cadence single dish monitoring observations at~1--3.5\,mm will help to constrain theoretical models of emission mechanisms as well the the site of gamma-ray production. Even more so,~3.5\,mm VLBI observations with the GMVA and~1.3\,mm VLBI observations with the EHT will help resolving the emission features and hence, foster the understanding of the structure and formation processes of AGN jets.

%%%%%%%%%%%%%%%%%%%%%%%%%%%%%%%%%%%%%%%%%%
%\section{Conclusions}
%
%This section is not mandatory, but can be added to the manuscript if the discussion is unusually long or complex.
%
%%%%%%%%%%%%%%%%%%%%%%%%%%%%%%%%%%%%%%%%%%
\vspace{6pt} 
%
%%%%%%%%%%%%%%%%%%%%%%%%%%%%%%%%%%%%%%%%%%
%% optional
%\supplementary{The following are available online at \linksupplementary{s1}, Figure S1: title, Table S1: title, Video S1: title.}
\authorcontributions{conceptualization, M.Ba. and H.F.;
%methodology, X.X.;
%software, X.X.;
%validation, X.X., Y.Y. and Z.Z.;
%formal analysis, X.X.;
%investigation, X.X.;
%resources, X.X.;
%data curation, X.X.;
writing—original draft preparation, M.Ba.;
writing—review and editing, M.Ba., M.Bö., and H.F.;
%visualization, X.X.;
%supervision, X.X.;
project administration, H.F.;
funding acquisition, H.F. and M.Bö.}
%For research articles with several authors, a short paragraph specifying their individual contributions must be provided. The following statements should be used “conceptualization, X.X. and Y.Y.; methodology, X.X.; software, X.X.; validation, X.X., Y.Y. and Z.Z.; formal analysis, X.X.; investigation, X.X.; resources, X.X.; data curation, X.X.; writing—original draft preparation, X.X.; writing—review and editing, X.X.; visualization, X.X.; supervision, X.X.; project administration, X.X.; funding acquisition, Y.Y.”, please turn to the  \href{http://img.mdpi.org/data/contributor-role-instruction.pdf}{CRediT taxonomy} for the term explanation. Authorship must be limited to those who have contributed substantially to the work reported.}
%
%%%%%%%%%%%%%%%%%%%%%%%%%%%%%%%%%%%%%%%%%%
\funding{This work is partly supported by the ERC Synergy Grant ``BlackHoleCam: Imaging the Event Horizon of Black Holes'' (Grant~610058) and the National Research Foundation~(NRF)\footnote{Any opinion, finding and conclusion or recommendation expressed in this material is that of the authors and the NRF does not accept any liability in this regard.} and the Department of Science and Technology of the Republic of South Africa through the South African Research Chair Initiative under SARChI grant~64798.}%Please add: ``This research received no external funding'' or ``This research was funded by NAME OF FUNDER grant number XXX.'' and  and ``The APC was funded by XXX''. Check carefully that the details given are accurate and use the standard spelling of funding agency names at \url{https://search.crossref.org/funding}, any errors may affect your future funding.}
%
%%%%%%%%%%%%%%%%%%%%%%%%%%%%%%%%%%%%%%%%%%
\acknowledgments{M.Ba. wants to thank Alan~Marscher and Svetlana~Jorstad for fruitful discussions.}
%
%%%%%%%%%%%%%%%%%%%%%%%%%%%%%%%%%%%%%%%%%%
\conflictsofinterest{The authors declare no conflict of interest.}%Declare conflicts of interest or state ``The authors declare no conflict of interest.'' Authors must identify and declare any personal circumstances or interest that may be perceived as inappropriately influencing the representation or interpretation of reported research results. Any role of the funders in the design of the study; in the collection, analyses or interpretation of data; in the writing of the manuscript, or in the decision to publish the results must be declared in this section. If there is no role, please state ``The funders had no role in the design of the study; in the collection, analyses, or interpretation of data; in the writing of the manuscript, or in the decision to publish the results''.} 
%
%%%%%%%%%%%%%%%%%%%%%%%%%%%%%%%%%%%%%%%%%%
% optional
\abbreviations{The following abbreviations are used in this manuscript (in order of appearance):\\
\noindent 
\begin{tabular}{@{}ll}
AGN	& Active Galactic Nucleus\\
SSC	&	Synchrotron Self-Compton\\
SED	& Spectral Energy Distribution\\
AMT	&	Africa Millimetre Telescope, \url{https://www.ru.nl/blackhole/africa-millimetre-telescope}\\
ALMA	& Atacama Large Millimeter/submillimeter Array, \url{https://www.almaobservatory.org}\\
SEST	&	Swedish ESO Submillimetre Telescope, \url{http://www.apex-telescope.org/sest}\\
EHT	& Event Horizon Telescope, \url{https://eventhorizontelescope.org}\\
MAXI	& Monitor of All-sky X-ray Image, aboard the International Space Station,\\
	& $\quad$\url{http://maxi.riken.jp/top}\\
\emph{Swift}-BAT	& Burst Alert Telescope, aboard the Neil Gehrels Swift Observatory,\\
	&$\quad$\url{https://swift.gsfc.nasa.gov}\\
\emph{AGILE}-GRID	& Gamma Ray Imaging Detector, aboard the Astro-Rivelatore Gamma a Immagini\\
	&$\quad$Leggero satellite, \url{http://agile.rm.iasf.cnr.it}\\
\emph{Fermi}-LAT	& Large Array Telescope, aboard the Fermi Gamma-ray Space Telescope,\\
	&$\quad$\url{https://fermi.gsfc.nasa.gov}\\
HAWC	& High-Altitude Water Cherenkov Observatory, \url{https://www.hawc-observatory.org}\\
SMARTS	& Small and Moderate Aperture Research Telescope System,\\
	&$\quad$\url{http://www.astro.yale.edu/smarts/glast}\\
GASP	&	GLAST-AGILE Support Programme,\\
	&$\quad$\url{http://www.oato.inaf.it/blazars/webt/gasp/homepage.html}\\
WEBT	&	Whole Earth Blazar Telescope, \url{http://www.oato.inaf.it/blazars/webt}\\
OVRO	&	Owens Valley Radio Observatory, \url{http://www.astro.caltech.edu/ovroblazars}\\
UMRAO	&	University of Michigan Radio Astronomy Observatory,\\
	&$\quad$\url{https://dept.astro.lsa.umich.edu/datasets/umrao.php}\\
IACT	& Imaging Atmospheric Cherenkov Telescope\\
VERITAS	& Very Energetic Radiation Imaging Telescope Array System,\\
	&$\quad$\url{https://veritas.sao.arizona.edu}\\
MAGIC	& Major Atmospheric Gamma Imaging Cherenkov, \url{https://magic.mpp.mpg.de}\\
H.E.S.S.	& High Energy Stereoscopic System, \url{https://www.mpi-hd.mpg.de/hfm/HESS}\\
KVA	& \url{http://users.utu.fi/kani/1m}\\
ATOM	& Automatic Telescope for Optical Monitoring,\\
	&$\quad$\url{https://www.lsw.uni-heidelberg.de/projects/hess/ATOM}\\
VLBI	& Very Long Baseline Interferometry\\
\emph{CGRO}-EGRET	& Energetic Gamma Ray Experiment Telescope, aboard the Compton Gamma Ray\\
	& $\quad$Observatory, \url{https://heasarc.gsfc.nasa.gov/docs/cgro/cgro/egret.html}\\
MOJAVE	& Monitoring Of Jets in Active galactic nuclei with VLBA Experiments,\\
	& $\quad$\url{http://www.physics.purdue.edu/MOJAVE}\\
TANAMI	& Tracking Active galactic Nuclei with Austral Milliarcsecond Interferometry,\\
	& $\quad$\url{http://pulsar.sternwarte.uni-erlangen.de/tanami}\\
F-GAMMA	& FERMI-GST AGN Multi-frequency Monitoring Alliance,\\
	& $\quad$\url{https://www3.mpifr-bonn.mpg.de/div/vlbi/fgamma/fgamma.html}\\
VLBA	& Very Long Baseline Array, \url{https://science.lbo.us/facilities/vlba}\\
VLBA-BU-BLAZAR	& VLBA Boston University Blazar monitoring project,\\
	&$\quad$\url{http://www.bu.edu/blazars/VLBAproject.html}\\
APEX	& Atacama Pathfinder EXperiment, \url{http://www.apex-telescope.org}\\
IRAM	& Institut de Radioastronomie Millim{\'e}trique, \url{http://iram-institute.org}\\
POLAMI	& POLarimetric monitoring of AGN at MIllimetre Wavelengths, \url{http://polami.iaa.es}\\
MARMOT	& Monitoring of $\gamma$-ray Active galactic nuclei with Radio, Millimetre and Optical\\
	&$\quad$Telescopes, \url{http://www.astro.caltech.edu/marmot}\\
GMVA	& Global mm-VLBI Array, \url{https://www3.mpifr-bonn.mpg.de/div/vlbi/globalmm}
\end{tabular}}
%
%%%%%%%%%%%%%%%%%%%%%%%%%%%%%%%%%%%%%%%%%%
%% optional
%\appendixtitles{no} %Leave argument "no" if all appendix headings stay EMPTY (then no dot is printed after "Appendix A"). If the appendix sections contain a heading then change the argument to "yes".
%\appendixsections{multiple} %Leave argument "multiple" if there are multiple sections. Then a counter is printed ("Appendix A"). If there is only one appendix section then change the argument to "one" and no counter is printed ("Appendix").
%\appendix
%\section{}
%\unskip
%\subsection{}
%The appendix is an optional section that can contain details and data supplemental to the main text. For example, explanations of experimental details that would disrupt the flow of the main text, but nonetheless remain crucial to understanding and reproducing the research shown; figures of replicates for experiments of which representative data is shown in the main text can be added here if brief, or as Supplementary data. Mathematical proofs of results not central to the paper can be added as an appendix.
%
%\section{}
%All appendix sections must be cited in the main text. In the appendixes, Figures, Tables, etc. should be labeled starting with `A', e.g., Figure A1, Figure A2, etc. 
% References, variant A: internal bibliography
%=====================================
\reftitle{References}
%\begin{thebibliography}{999}
% Reference 1
%\bibitem[Author1(year)]{ref-journal}
%Author1, T. The title of the cited article. {\em Journal Abbreviation} {\bf 2008}, {\em 10}, 142-149, doi:xxxxx.
% Reference 2
%\bibitem[Author2(year)]{ref-book}
%Author2, L. The title of the cited contribution. In {\em The Book Title}; Editor1, F., Editor2, A., Eds.; Publishing House: City, Country, 2007; pp. 32-58, ISBN.
%\end{thebibliography}
%
% The following MDPI journals use author-date citation: Arts, Econometrics, Economies, Genealogy, Humanities, IJFS, JRFM, Laws, Religions, Risks, Social Sciences. For those journals, please follow the formatting guidelines on http://www.mdpi.com/authors/references
% To cite two works by the same author: \citeauthor{ref-journal-1a} (\citeyear{ref-journal-1a}, \citeyear{ref-journal-1b}). This produces: Whittaker (1967, 1975)
% To cite two works by the same author with specific pages: \citeauthor{ref-journal-3a} (\citeyear{ref-journal-3a}, p. 328; \citeyear{ref-journal-3b}, p.475). This produces: Wong (1999, p. 328; 2000, p. 475)
%
%=====================================
% References, variant B: external bibliography
%=====================================
\externalbibliography{yes}
\bibliography{AMT-AGNmonitoring}
%
%%%%%%%%%%%%%%%%%%%%%%%%%%%%%%%%%%%%%%%%%%
%% optional
%\sampleavailability{Samples of the compounds ...... are available from the authors.}
%
%% for journal Sci
%\reviewreports{\\
%Reviewer 1 comments and authors’ response\\
%Reviewer 2 comments and authors’ response\\
%Reviewer 3 comments and authors’ response
%}
%
%%%%%%%%%%%%%%%%%%%%%%%%%%%%%%%%%%%%%%%%%%
\end{document}